\documentclass[twocolumn,prc,aps,floatfix]{revtex4}
\usepackage[dvips]{graphicx}
\begin{document}
\title{Comment on ``Time modulation of the K-shell electron capture decay rates of H-like heavy ions at GSI experiments ''}
\author{V.V. Flambaum}
\affiliation{
 School of Physics, The University of New South Wales, Sydney NSW
2052, Australia and ECT*, Villa Tambosi, I-38100 Villazzano  (Trento), Italy
}
\date{\today}
\begin{abstract}
According to  GSI experiment \cite{GSI}, the rate of the daughter ion production by the
K-shell electron capture (with neutrino emission) in H-like ions $^{142}Pm$, $^{140}Pr$ and $^{152}I$ is modulated with
a period of 6 to 7 seconds. In the Letter \cite{Ivanov} (also \cite{Lipkin}) there is a  claim that
neutrino oscillations may explain this result. In Ref. \cite{Lambiase} a different explanation
is suggested which is based on the rotation of electron and nuclear spins. The aim of this Comment is to
show why these explanations \cite{Ivanov,Lipkin,Lambiase} are not satisfactory and to discuss other possibilities.
\end{abstract}
\maketitle

    To treat this problem it is convenient to use the basis of neutrino mass eigenstates in free space,
    $\nu_1$, $\nu_2$ and $\nu_3$. In this basis we have three different reactions
$e$ + $^{142}Pm$ $\to$ $\nu_1$ + $^{142}Nd$, $e$ + $^{142}Pm$ $\to$ $\nu_2$ + $^{142}Nd$ and
     $e$ + $^{142}Pm$ $\to$ $\nu_3$ + $^{142}Nd$ (the amplitudes of these reactions are proportional to the expansion coefficients $U_{e1}$, $U_{e2}$ and $U_{e3}$ correspondingly where
     $\nu_e=U_{e1} \nu_1 +U_{e2}\nu_2+U_{e3}\nu_3$). The final states of these reactions are different
     and orthogonal to each other since $\nu_1$, $\nu_2$  and $\nu_3$ are different particles.
     Therefore, the amplitudes of these reactions can not interfere, and no oscillations are possible.
     The situation here is similar, for example, to decays like $\pi$ $\to$ $\mu \nu$ and $\pi$ $\to$ $e\nu$, where
     we have different particles in the final states of two reactions and no oscillations in the $\pi$-meson decay.

  Why neutrino oscillations appear in other reactions? Weak interaction used to detect the neutrino projects
$\nu_1$, $\nu_2$  and $\nu_3$ to the state $\nu_e$ (if electron is produced). As a result, we have three amplitudes (with intermediate
states $\nu_1$, $\nu_2$  and $\nu_3$) which lead to the same  state $\nu_e$. Here we have interference
and oscillations.  In the case of the GSI experiment this does not happen.

   One may speculate about a different possibility to produce the interference. Neutrinos pass through the strong
electromagnetic field. If neutrinos have magnetic moments which are not diagonal in the $\nu_1$, $\nu_2$, $\nu_3$
basis (for example, if the magnetic moments are diagonal in the  $\nu_e$, $\nu_{\mu}$  and $\nu_{\tau}$ basis)
one, in principle,  may have three amplitudes leading to the same final state and neutrino oscillations.
However, it is hard to link this possibility to the GSI oscillations.

    In Ref. \cite{Lambiase} the GSI oscillations were explained by oscillations between the ionic states with different  total angular momentum ${\bf F}={\bf I}+{\bf s}$ which is due to the rotation of the electron spin
     ${\bf s}$ and the nuclear spin ${\bf I}$ in the magnetic field $B$ of the storage ring in the GSI experiment. For example, the  $^{142}Pm$ nucleus has spin $I=1$, and we have two values of $F$ possible, $F=3/2$ and $F=1/2$. If probability of the weak electron capture depends on $F$, the oscillations between the states    $F=3/2$ and $F=1/2$
    would lead to the oscillations in the electron capture rate. In principle, such an effect may  exist, however,
    the authors of \cite{Lambiase} overestimated it by many orders of magnitude; also, the frequency of such oscillations would be  $\sim 10^{14}$ Hz instead of $\sim 10^{-1}$  Hz. Indeed,the authors of Ref.  \cite{Lambiase} neglected the hyperfine interaction between the electron spin and nuclear spin. However, in heavy hydrogen-like ions the hyperfine interaction is enhanced by a factor $Z^3 \sim 10^5$ times, where $Z$ is the proton number. This interaction exceeds the interaction of the electron
and nuclear spins with the magnetic field by many orders of magnitude and determines the interval between the hyperfine levels $E_{Hf}$. Mixing of the hyperfine states $F=3/2$ and $F=1/2$ is proportional to $\mu_B B/ E_{Hf}$ and very small (here $\mu_B$ is the Bohr magneton).
Moreover, this mixing will not lead to the oscillations if the switching of the field in the ion reference frame is slow (adiabatic) in comparison
with the hyperfine period $h/E_{Hf} \sim 10^{-14}$ s. Finally, the oscillations, if any, will be determined by the energy difference between the hyperfine states and have a frequency $E_{Hf}/h \sim 10^{14}$ Hz.

The motion
of the ion (on the circular orbit in the ring) does not change the value of $F$, it only changes the evolution of the direction of ${\bf } F$ which, however, does not influence the electron capture rate.

  Note that any link between the GSI oscillations and atomic phenomena would  not look natural. Indeed,
   the energy intervals and interactions in the ion including the interactions with external fields in the
   storage ring in the  GSI experiment
    exceed 0.1 Hz by many orders of magnitude. Such phenomena like  Rabi oscillations with a noticeable modulation amplitude would require a special experimental arrangement and do not happen by chance.

I am grateful to John Schiffer who asked me to consider this problem and to Achim Richter
who turned my attention to the article \cite{Ivanov} and provided many useful comments. I am also
grateful to ECT* staff for the support and hospitality.


\begin{thebibliography}{99}
\bibitem{GSI} Yu. A. Litvinov et al., Phys. Rev. Lett.{\bf 99}, 262501 (2007); Phys.Lett.B{\bf 664}, 162 (2008).
\bibitem{Ivanov} A.N. Ivanov and P. Kienle,  Phys. Rev. Lett.{\bf 103}, 062502 (2009).
\bibitem{Lipkin} H.J. Lipkin, arxiv: 0905.1216, 0904.4913, 0805.0435, 0801.1465 .
\bibitem{Lambiase} G. Lambiase, G. Papini, G. Scarpetta, arxiv: 0811.2302 .
\end{thebibliography}
\end{document}